\documentclass[aps,prd,onecolumn,showpacs,showkeys,superscriptaddress]{revtex4-2}
\usepackage{latexsym}
\usepackage{amssymb}
\usepackage{amsmath}
\usepackage{amscd}
\usepackage{amsthm}
\usepackage{graphicx}
\usepackage{textcomp}
\usepackage{colortbl}
\usepackage[colorlinks]{hyperref}
\usepackage[font={footnotesize,it}]{caption}
\usepackage{multirow}
\usepackage[T1]{fontenc}
\usepackage{ae,aecompl}
\usepackage{subcaption}

\begin{document}
\title{Cosmological Implications of the Gong–Zhang Parameterization in Rastall Gravity: A Deep Learning and Observational Study}
\vspace{20mm}

\author{Vinod Kumar Bhardwaj}
\email{dr.vinodbhardwaj@gmail.com}
\affiliation{Department of Mathematics, GLA University, Mathura-281 406, Uttar Pradesh, India}

\author{Anil Kumar Yadav}
\email[]{abanilyadav@yahoo.co.in}
\affiliation{Department of Physics, United College of Engineering and Research, Greater Noida - 201310, India}

\author{Manish Kalra}
\email{manishkalra2012@gmail.com}
\affiliation{Department of Electronics and Communication Engineering, GLA University, Mathura-281 406, Uttar Pradesh, India}

\author{Pankaj}
\email{pankaj.fellow@yahoo.co.in}
\affiliation{School of Sciences, IILM University, Knowledge Park-II, Greater Noida - 201310, India}

\author{Rajendra Prasad}
\email{drrpnishad@gmail.com}
\affiliation{Department of Physics, Galgotias College of Engineering and Technology, Greater Noida, 201310 India}

\vspace{2cm}


\begin{abstract}
In this study, we have explored the cosmological dynamics of an isotropic, homogeneous universe in Rastall gravity. For this purpose, we use the parameterization of the EoS parameter in the form $\omega(z) = \frac{\omega_{0}}{(z+1)} $ to derive the explicit solution of the field equations in Rastall gravity. We constrained the cosmological parameters for the derived model by the Markov Chain Monte Carlo (MCMC) approach utilizing OHD, BAO, and Pantheon plus compilation of SN Ia datasets. We also constrained the model parameters using deep learning techniques and the CoLFI Python package. This paper introduces an innovative deep-learning approach for parameter inference. The deep learning method significantly surpasses the MCMC technique regarding optimal fit values, parameter uncertainties, and relationships among parameters. This conclusion is drawn from a comparative analysis of the two methodologies. Additionally, we determined the transition redshift $z_t = 0.941$, which signifies the shift in the Universe's model from an early deceleration phase to the present acceleration phase. The diagnosis of the model with diagnostic tools like statefinders, jerk parameter, and $O_m$ diagnostics are presented and analyzed. The validation of the model's energy conditions is also examined.
\end{abstract}

\maketitle 
\section{Introduction}\label{sec1}
The experimental findings like $H(z)$ data, Ia Supernovae \cite{ref1,ref2}, Wilkinson Microwave Anisotropy Probe (WMAP) \cite{ref3,ref4}, Baryon Acoustic Oscillations (BAO) \cite{ref5}, Cosmic Microwave Background (CMB) \cite{ref6}, and Large Scale Structure (LSS) \cite{ref7} confirmed the present era of accelerating universe. Although the cosmological models based on GR show an agreement with various observational findings \cite{ref1,ref2,ref3}, still, the observed accelerated expansion of present cosmos argue the validity of General Relativity (GR) on large scales \cite{ref8}. An exotic cosmic fluid with large negative pressure having a mysterious kind of energy density (known as dark energy (DE)) is postulated to claim this observed acceleration. Because of its repulsive nature, the cosmological constant makes it an appropriate substitute for DE \cite{ref8,ref9}. However, cosmic models with a cosmological constant face the complications of cosmic fluke and fine-tuning. Apart from these issues, the $\Lambda$CDM model also suffers from $H_0$ tension. To address these issues and to describe the origin and nature of dark energy, several dynamical dark energy models, including modified gravity models, models based on extra dimensions, and scalar field models, have been proposed in the literature \cite{ref10,ref11,ref12,ref13,ref14,ref15,ref16}.\\

The principle of equivalence is a cornerstone of Einstein's theory of general relativity. Nevertheless, there are instances where this rule is violated, resulting in modifications of gravity that deviate from GR anticipations. The divergence-free tensor, which exhibits minimal correlation with the geometry of spacetime, has been utilized to establish an energy-momentum source in General Relativity and its variations in modified gravitational theories \cite{ref17,ref18}. However, it is argued that the process of particle formation violates the property of the energy-momentum tensor (EMT), which leads to the energy-momentum conservation equation ($\nabla_{\nu}^{\mu \nu} = 0$) \cite{ref19,ref20,ref21,ref22}. Therefore, it is permissible to disregard the EMT conservation requirement in favor of a different gravitational theory. In the present scenario, the utilization of modified gravity theories \cite{ref23,ref24,ref25} is both advantageous and attractive in tackling the significant challenges associated with the typical cosmological models, including `dark energy (DE)' \cite{ref1,ref26} and `dark matter (DM)' \cite{ref27,ref28}. It is essential to recognize that any alternative to General Relativity must be valid. Numerous researchers have extensively explored modified theories of gravity \cite{ref29,ref30,ref31,ref32,ref33,ref34,ref35}.\\

Rastall's gravity proposed by Peter Rastall in 1972 \cite{ref36} is one of the interesting theory among the alternative gravitational theories. The law of conservation of energy-momentum tensor (EMT) is analyzed under circumstances, such as in Minkowski flat space-time or specifically in weak gravitational arenas in Rastall's framework. This raises questions regarding its relevance in curved space-time. The postulate that the covariant derivative of the energy-momentum tensor (EMT) is zero is no longer valid in Rastall's theory. Rather, we examine a vector field that corresponds to the gradient of the Ricci scalar, represented as $\nabla_{\nu} T^{\mu \nu} \propto \nabla^{\mu}R$. In Rastall’s framework, the curvature is coupled with the violation of the energy and momentum conservation laws, and it can be interpreted as a conventional representation of quantum phenomena \cite{ref37}. Additionally, the curvature-matter theory of gravity defines a non-minimal interaction between matter and geometry, resulting in the failure of the traditional conservation law for energy-momentum \cite{ref38,ref39,ref40,ref41}.\\ 

Recently, significant advancements have been achieved in the investigation of various aspects of Rastall gravity including the role of the Rastall constraints in DE fluctuations \cite{ref42}, and its relationship with the `Brans-Dicke scalar field’ \cite{ref43}, and the structure of neutron stars \cite{ref44}. Researchers have also examined various cosmic eras within this theoretical framework offering an extensive analysis of both the theoretic and experimental dimensions of Rastall gravity \cite{ref45,ref46,ref47,ref48,ref49}. Remarkably, Rastall gravity is unaffected by issues related with age and entropy \cite{ref50}, and it provides clarifications for both accelerated expansion scenario and inflationary eras \cite{ref45,ref51}. Moradpour et al.\cite{ref51}, suggest that a non-minimal coupling between pressure-less matter and geometry within Rastall gravity may mimic the effects of dark energy, which could explain the current phase of cosmic acceleration. The Rastall gravity framework has been rigorously analyzed throughout the cosmic era, encompassing early inflation, the matter-dominated era, and the phase of accelerated expansion \cite{ref52}.Additionally, the cosmic model of Rastall gravity introduced in \cite{ref53} is compatible with the $\Lambda$CDM model in the later stages of the cosmos. In Ref. \cite{ref52}, it has been investigated that a Rastall parameter also serve as a viable solution to the initial singularity issue. Therefore, as a final remark, we emphasise that the theoretical appeal of Rastall gravity resides in its minimalist yet covariant extension of general relativity, achieved by generalizing the matter-geometry coupling 
through a relaxed conservation law. From the phenomenological standpoint, it offers a promising framework for explaining cosmic acceleration\cite{Fabris2015,Akarsu2020}, unifying dark sector dynamics \cite{Mehdizadeh2018}, and remaining consistent with current observational data \cite{Tang2019}, while also being interpretable as an effective manifestation of quantum effects in curved spacetime \cite{Moradpour2017}. In comparison with other modified gravity  approaches such as $f(R)$ gravity \cite{DeFelice2010}, $f(T)$ gravity \cite{Cai2016}, and scalar-tensor theories (e.g., 
Brans-Dicke) \cite{Brans1961}, Rastall gravity stands out for its simplicity, mathematical economy, and observational viability, while still accommodating subtle deviations from general relativity \cite{Visser2018}. Furthermore, it is worthwhile to comment on the physical status of Rastall gravity in light of the Ref. \cite{Visser2018} in which, author has investigated that Rastall gravity can be mathematically recast as General Relativity with a modified effective energy momentum tensor. We acknowledge this equivalence at the level of field equations. However, our goal is not to introduce new gravitational degrees of freedom, but to explore the cosmological consequences of a non conserved matter part. When phenomenological dark energy parameterizations such as the Gong Zhang model are imposed on the physical matter sector, the resulting cosmological dynamics differ from those of GR because the continuity equation is modified. Consequently, the expansion history, effective equation of state, and transition redshift become Rastall dependent. In this sense, our work probes the observational viability of non conservative cosmological scenarios rather than a fundamentally new theory of gravity. \\

By employing modern Deep Learning approaches, the aim was to analyze the accelerated expansion of the universe while simultaneously estimating cosmological parameters. Utilizing CoLFI python package \cite{ref54}, the cosmological parameters are estimated using Artificial Neural Networks (ANN), Mixture Density Networks (MDN), and Mixture of Gaussians (MNN) techniques \cite{ref55,ref56,ref57,ref58}. This advancement significantly facilitated the understanding of conditional probability densities derived from observational data and posterior distributions \cite{ref54,ref55,ref56}. To enhance the efficiency of neural network training and improve the accuracy of parameter predictions, hyper ellipsoid parameters were introduced \cite{ref57,ref58}. A comparative study between our neural network techniques and the conventional Markov Chain Monte Carlo (MCMC) approach revealed that MNN produced results nearly identical to those of MCMC, underscoring its effectiveness and reliability.\\

The proposed model of the universe accounts for late-time acceleration without the need for dark energy and avoids issues associated with the cosmological constant. Therefore, it is essential to consider current observational data when simulating Rastall gravity. We have shown that neural network methods can effectively estimate cosmological parameters, serving as a practical alternative to MCMC techniques. The Rastall gravity framework facilitates the exploration of various gravity theories and their influence on cosmic expansion through the use of ANN, MDN, and MNN methodologies. This research is part of a broader study integrating machine learning with cosmology, demonstrating the capability of machine learning to address intricate cosmic dynamics and expansion challenges. Our evaluation of density, cosmic pressure, the equation of state, and deceleration parameters lends credibility to our findings. The given paper is organized in the following manner. A solution of Rastall gravity field equations using EsS parameterization is mentioned in Section II. The observational constraints on model parameters using the MCMC approach and deep learning utilizing OHD, BAO, and Pantheon datasets are presented in Section III. Some cosmological features of the model, like cosmic pressure, energy density, and deceleration parameter, along with diagnostic tools like statefinders, jerk parameter, and $O_m$ diagnostic are presented and analyzed in Section IV. The energy conditions on the viability of the model are discussed in Section V. The information criteria and model selection is described in Section VI.  Finally the summary of this work is presented in Section VII.\\

\section{Metric and the Field Equations}
The modified theory based on Rastall's gravity is one of the intriguing subclasses among the alternative theories of gravity\cite{ref36}. Our goal is to extract solutions, carefully taking into account both the source of the matter and the gravitational background. The Bianchi identities remain valid in Rastall's gravity, but the conservation of the stress-energy tensor of the gravity source is violated (since $T^{i j}_{; i} = \lambda R^{; j}$, where $R$ is the Ricci scalar and $\lambda$ is the Rastall parameter). The covariant conservation equation of the energy-momentum tensor in this GR modification is expressed as $\nabla_{j} T^{i j} = 0$ and can be expressed in more generalized form as $\nabla_{j} T^{i j} =  u^{i}$\cite{ref36,ref46,ref59}, here to retrieve this again in GR, the right side should be zero. Consequently, vector $u^{i}$ is considered as $u^{i}=\lambda \nabla^{i}R$. The field equations in the framework of Rastall gravity can be expressed as\cite{ref36,ref59}
\begin{equation}\label{eq1}
R_{ij} -\frac{1}{2} (1-2 k \lambda) g_{ij} R = k T_{i j}
\end{equation}
where $R_{i j}$ is the Ricci-tensor, $g_{ij}$ is metric-tensor, and $T_{ij}$ is the energy-momentum tensor respectively. $k$ denotes the gravitational constant, which is to be determined in consistent with Newtonian limit. The Einstein field equations can be retrieve for $\lambda=0$ and $k = 8 \pi$ whenever $T^{i j}_{; i} = 0$. Then, the trace of Eq. (\ref{eq1}) can be recasts as\cite{ref59}
\begin{equation}\label{eq2}
(4 k \lambda - 1) R = k T
\end{equation}
For either $R = 0$ or $ k\lambda = \frac{1}{4}$, we get a trace-less energy-momentum tensor ($T = 0$). Rastall gravity has similar characteristics to GR for $R = 0$, while the trace of energy-momentum reduces to zero for $ k\lambda = \frac{1}{4}$.
For $ k\lambda \neq \frac{1}{4}$, the field equation with the cosmological constant $\Lambda$ is given by
\begin{equation}\label{eq3}
G_{ij} +\Lambda g_{ij}+ k \lambda g_{ij} R = k T_{i j}
\end{equation}
where $G_{ij}$ denotes the standard Einstein tensor.
The Friedmann-Lema\^{i}tre-Robertson-Walker (FLRW) space-time metric for a homogeneous and isotropic universe is considered as
\begin{equation}\label{eq4}
ds^{2} = -dt^2+a^{2}(t) \bigg(\frac{dr^2}{1-K r^2}+r^2 d\Omega^2\bigg).
\end{equation}
where $d\Omega^2 = d\theta^2+sin^{2}\theta d\phi^2$; $a(t)$ denotes the time-dependent scale factor; and spatial parameter $K$ represents the open, flat, and closed space sections for the values $-1$, $0$, and $+1$ respectively.\\
For the perfect fluid, the field equations in the Rastall gravity are expressed as \cite{ref36,ref46,ref47,ref59}
\begin{equation}\label{eq5}
3 (1-4 k \lambda) H^2 - 6 \lambda  k \dot{H} - 3 (2 \lambda k-1) K a^{-2} = k \rho ,
\end{equation}
\begin{equation}\label{eq6}
- 3 (4\lambda k - 1) H^2 +2(1- 3\lambda k) \dot{H} -  (6 \lambda k -1)K a^{-2} = - k p.
\end{equation}
where $p$ and $\rho$ respectively, denote the cosmic pressure and energy density, and over dot denotes the derivative with respect to time $t$. The scale factor $a$ and Hubble parameter $H$ are related as $H = \frac{\dot{a}}{a}$. The isotropic pressure for the model is connected with energy density by relation $p = \omega \rho$ called as `equation of state (EoS)'. By applying the Bianchi identity ($G^{; j}_{ij} = 0$, the continuity equation is derived as \cite{ref59}
\begin{equation}  \label{eq7}
(3 k \lambda-1) \dot{\rho}+3 k \lambda \dot{p}+3 (4 k \lambda-1) H (\rho+p) = 0.
\end{equation}
Since most of cosmological observational datasets are exist in terms of redshift $z$ rather than cosmic time $t$. The scale factor $a(t)$ is connected with redshift $z$ using the relation $\frac{a_0}{a} = (1+z)$, where $a_0$ is the present value of scale factor,  which is considered as $1$ in the present analysis. Using this relation we also get $\dot{H(t)} = -(1+z) H(z) H'(z)$, where $H'(z) = \frac{dH}{dz}$. In terms of redshift $z$, the field equations  (\ref{eq5}) and  (\ref{eq6}) can be recast as
\begin{eqnarray}\label{eq8}
& 3 (1-4 \lambda k) H^2 + 6  \lambda k (z+1) H H'(z) \nonumber\\
&- 3  K  (2 \lambda k - 1) (z+1)^2 = k \rho ,
\end{eqnarray}
\begin{eqnarray}\label{eq9}
&3 (1-4 \lambda k) H^2 + 2(3\lambda k -1)  (z+1) H H'(z)\nonumber\\
&+ (1-6 \lambda k )  (z+1)^2 K = - k p.
\end{eqnarray}
For flat universe ($K = 0$), using EOS relation ($p = \omega \rho$), from equations  (\ref{eq8}) and  (\ref{eq9}), we get the expression
\begin{eqnarray}\label{eq10}
3 (1-4 \lambda) [1+\omega(z)] H^2(z) -[2-6 \lambda (1+\omega(z))]  (1+z) H(z) \frac{d H}{dz} = 0.
\end{eqnarray}
where for numerical simplification $k$ is considered as 1. Since above expression consist of two variables $\omega$ and $H(z)$, thus to get explicit solution we consider the redshift parameterization of EoS parameter $\omega(z)$ proposed by Gong-Zhang \cite{ref60} in th form 
\begin{eqnarray}\label{eq11}
\omega(z) = \frac{\omega_0}{(1+z)}
\end{eqnarray}
with $\omega_0$ as the present value of the EoS parameter. Using the Gong-Zhang parametrization \cite{ref60} for the EoS of dark energy (DE), one can efficiently describe distinct cosmic scenarios of an expanding universe. In the early time, as the $z$ approaches to infinity, EoS parameter tends to zero ($\omega \sim 0$) that represents the big-bang scenario of universe. At $z=0$, the present value of the EoS parameter is $\omega_0$. At $z \to -1$, the EoS parameter approaches negative infinity, which indicates towards the dark energy dominated scenario in the remote future.  \\

\noindent Moreover, to test model robustness with other alternative EoS parametrizations, we confront the parametrization $\omega(z)=\omega_0/(1+z)$ with widely used alternatives, including constant $\omega(z)=\omega_0$, the Chevallier--Polarski--Linder (CPL) form $\omega(z)=\omega_0+\omega_a z/(1+z)$ \cite{Chevallier2001,Linder2003}, the Jassal--Bagla--Padmanabhan (JBP) form 
$\omega(z)=\omega_0+\omega_a z/(1+z)^2$ \cite{JBP2005}, the Barboza--Alcaniz (BA) parametrization \cite{Barboza2008}, 
and logarithmic models $\omega(z)=\omega_0+\omega_a \ln(1+z)$ \cite{Efstathiou1999}, as well as nonparametric Gaussian-process reconstructions \cite{Seikel2012}. Our choice has the advantage of analytic tractability, suppresses early dark energy ($\omega\to 0$ as $z\to\infty$), and smoothly recovers $\omega_0$ at $z=0$. In contrast, CPL allows for arbitrary high-$z$ limits, JBP returns to $\omega_0$ at early times, BA is bounded and smooth for all $z$, and the logarithmic form varies slowly but can diverge at high redshift. Within Rastall gravity, where nonconservation can mimic matter-like behaviour, degeneracies between $\omega_0$, $\Omega_{\rm de,0}$, and the Rastall parameter $\lambda$ 
must be carefully constrained with CMB \cite{Planck2018}, BAO, SN, and growth data. This comparative analysis ensures that apparent late-time acceleration is not an artifact of the chosen parametrization but a genuine prediction of the model 
\cite{Visser2018,Akarsu2020,Fabris2015IJMPD}.\\
 
\noindent From equations (\ref{eq10}) and  (\ref{eq11}), the expression of the Hubble parameter can be derived as
\begin{eqnarray}\label{eq12}
H(z) = H_0 \bigg[(z+1)^{3 \lambda -1} \left(\frac{(3 \lambda -1) z}{3 \lambda  (\omega_0+1)-1}+1\right)\bigg]^{\frac{4 \lambda -1}{2 \lambda  (3 \lambda -1)}}
\end{eqnarray}
Here, $H_0$ is the present value of Hubble parameter.\\ 

\noindent It is important to note that the General Relativity (GR) limit is properly recovered when the Rastall parameter $\lambda \rightarrow 0$. Therefore, to verify the GR limit, we study the behaviour of the Hubble parameter as $\lambda \to 0$. In this limit, the factors inside the bracket of Eq. (\ref{eq12}) approach unity, while the exponent diverges, leading to an indeterminate form $1^{\infty}$.\\ 
By expanding the logarithm of $H(z)$ to first order in $\lambda$, we obtain
\[
\ln\!\left(\frac{H(z)}{H_0}\right) \rightarrow \frac{3}{2}\ln(1+z).
\]
Exponentiation yields
\[
\lim_{\lambda\to0} H(z)=H_0(1+z)^{3/2},
\]
which corresponds to the standard GR result for a pressureless matter dominated Universe. This confirms that the model consistently reduces to General Relativity in the $\lambda\rightarrow 0$ limit.


\section{Data \& Methodology}
\noindent In recent years, Bayesian inference has increasingly been utilized for parameter estimation and model comparison in the field of cosmological research. In this section, our goal is to constrain the free parameters of the model using observational data sets and the $\chi^2$-minimization method. To find the values of $\omega_0$, $\lambda$, and $H_{0}$ in the model, we utilize the data sets listed below: \\

\noindent \textbf{OHD}: For constraining the model parameters, we have employed the 77 uncorrelated Hubble observations $H(z)$ that lie in the redshift range $0 \leq z \leq 2.36$. These $77$ $H(z)$ data points with its original references are compiles in Ref. \cite{Singh/2024Universe}. The $\chi^{2}$ function to comprise the covariance matrix is read as:
\[
\chi^2_{\mathrm{CC}} = (\mathbf{H}_{\text{obs}} - \mathbf{H}_{\text{model}})^T \cdot \mathbf{C}^{-1} \cdot (\mathbf{H}_{\text{obs}} - \mathbf{H}_{\text{model}}),
\]
where $\mathbf{H}_{\text{obs}}$ and $\mathbf{H}_{\text{model}}$ denote the observed and theoretical $H(z)$ values, respectively, and $\mathbf{C}$ represents the covariance matrix.\\

\noindent\textbf{Pantheon plus}: We utilize here the recent SN Ia observational data sets in the redshift range $0.001 < z < 2.26$ \cite{ref81} to constraint the model parameters. The Pantheon+ release presents 1701 light curves corresponding to $\approx 1550$ spectroscopically confirmed SNe Ia used in cosmological analyses\\ 

\noindent \textbf{BAO data}: We also considered the six BAO data points \cite{ref82,ref83,ref84}. For the BAO sample, the predictions from a sample of Galaxy Surveys like SDSS DR7 and 6dF, and WiggleZ have been utilized\cite{ref82,ref83,ref84}. The angular diameter distance for the sample is defined as $D_{A}=\frac{D_{L}}{(1+z)2}$, where $D_{L}$ indicates the proper angular diameter distance \cite{ref85}, and the dilation scale is described by $D_{V}(z)=\left[ D^{2}_{L}(z)*(1+z)^2*\frac{c \,  z}{H(z)} \right]^{1/3}$. 
\begin{table}
	\begin{center}
		\begin{tabular}{|c|c|c|c|c|c|c|}
			\hline
			$z_{BAO}$& $0.106$ & $0.2$	 & $0.35$  & $0.44$ & $0.6$ &$ 0.73$  \\
			\hline
			\small  $\varUpsilon(z)$ & \small$30.95 \pm 1.46$ & \small$17.55 \pm 0.60$  & \small$10.11 \pm 0.37$ & \small $8.44 \pm 0.67$ &\small $6.69 \pm 0.33$ & $ 5.45 \pm 0.31$\\
			\hline
		\end{tabular}\caption{ The values of $\varUpsilon(z)$ for different points of $z_{BAO}$ }
	\end{center}
\end{table}
\noindent Here, $\varUpsilon(z)= d_A(z_{*})/D_V(z_{BAO})$ and $z_{*}\approx 1091$.\\
For limiting the parameters of the model,  the chi-square estimator for the BAO sample is described in the following form \cite{ref85,ref86,ref87,ref88}.
\begin{equation}
	\chi^2_{BAO} = X^{T} C^{-1} X
\end{equation}
where 
\begin{align*}
	X &= \begin{pmatrix}
		\frac{d_A(z_{*})}{D_V(0.106)}-30.95\\           
		\frac{d_A(z_{*})}{D_V(0.20)}-17.55\\
		\frac{d_A(z_{*})}{D_V(0.35)}-10.11\\
		\frac{d_A(z_{*})}{D_V(0.44)}-8.44\\
		\frac{d_A(z_{*})}{D_V(0.60)}-6.69\\
		\frac{d_A(z_{*})}{D_V(0.73)}-5.45
	\end{pmatrix}
\end{align*}

\noindent The model parameters  $\omega_0$, $\lambda$, and $H_{0}$ of this model can be assessed using observational data and statistical methods. $E_{obs}$ represents the value of any observational data, while $E_{th}$ represents the values that have been theoretically estimated in the model of the Universe. By using statistical methods to compare the two values $E_{obs}$ and $E_{th}$, the model's parameters can be determined. To obtain the most accurate estimates, we utilized the $\chi^2$ estimator. If the standard error in the observed values is denoted by $\sigma$, the formula for the $\chi^2$ estimator is expressed as.
\begin{equation}
	\label{chi1}
	\chi^{2} = \sum_{i=1}^{N}\left[\frac{E_{th}(z_{i})- E_{obs}(z_{i})}{\sigma_{i}}\right]^{2}
\end{equation}
where, $E_{th}(z_{i})$ represents the theoretical values of the parameter, while $E_{obs}(z_{i})$ represents the observed values. $\sigma_{i}$ is the error and N is the number of data points. We obtain the best-fit parameter values for the derived model by minimizing the $\chi^2$ statistic.


\begin{figure}
\centering
\includegraphics[scale=0.7]{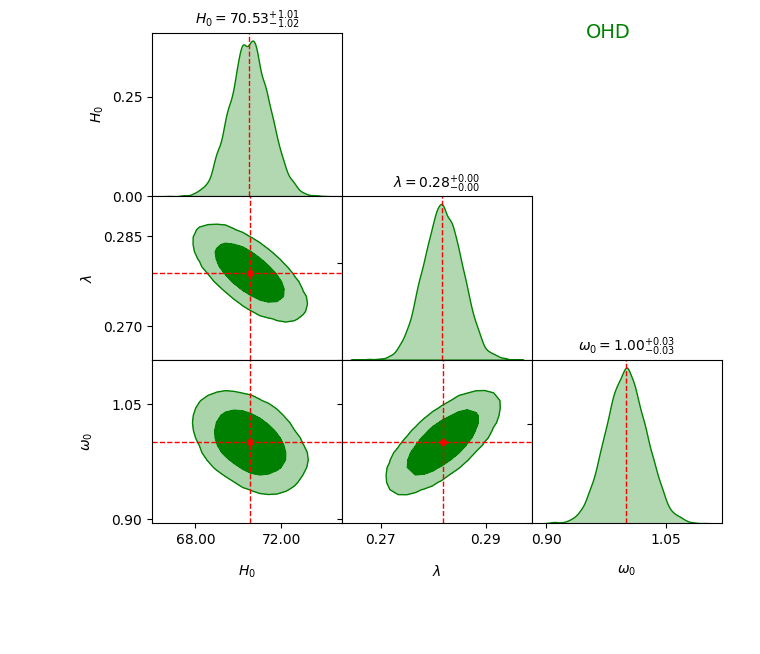}
\caption{One-dimensional marginalized distribution, and $2D$ contours with $1\sigma$ and $2\sigma$ confidence levels for the derived model of the Universe with 77 OHD datasets.}
\end{figure}
\begin{figure}
\centering
\includegraphics[scale=0.7]{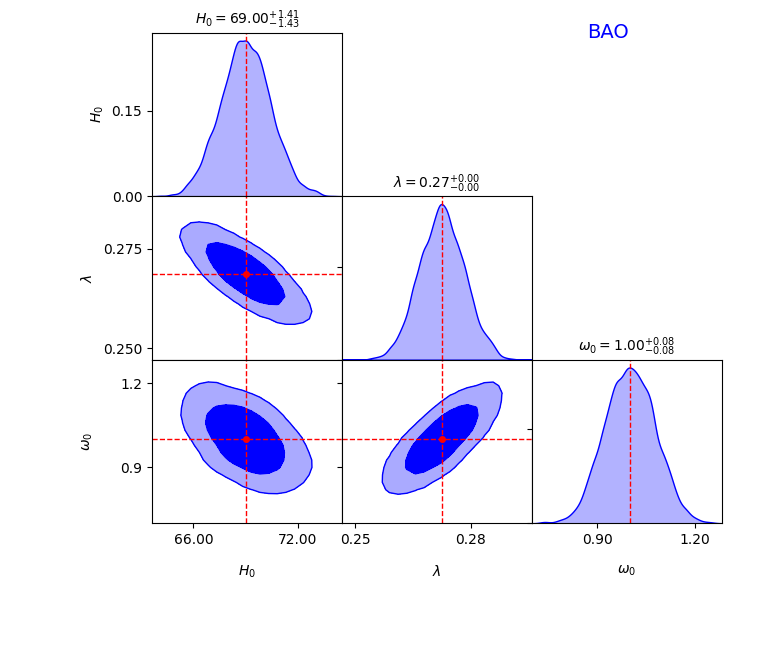}
\caption{One-dimensional marginalized distribution, and $2D$ contours with $1\sigma$ and $2\sigma$ confidence levels for the derived model of the Universe with BAO datasets.}
\end{figure}
\begin{figure}
\centering
\includegraphics[scale=0.7]{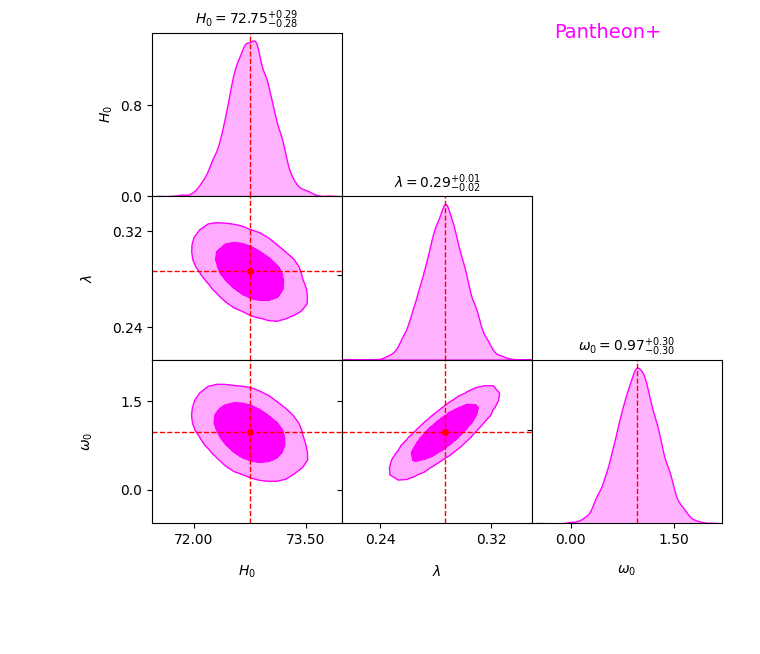}
\caption{One-dimensional marginalized distribution, and $2D$ contours with $1\sigma$ and $2\sigma$ confidence levels for the derived model of the Universe with Pantheon plus compilation of SN Ia datasets.}
\end{figure}
\section*{Neural Networks}
Neural networks become very relevant when computing advances led to the emergence of a new scientific field called deep learning, which is only concerned with the study of artificial neural networks. Currently, there are various kinds of neural networks—such as recurrent networks commonly applied to time series, convolutional neural networks excelling in image processing, auto-encoders employed for image denoising, and contemporary generative adversarial networks. In this study, we concentrate on the fundamental deep neural network, commonly referred to as the multi-layer perceptron (MLP). To estimate the parameters $(H_0,\lambda,\omega_0)$ from the OHD dataset, we used a supervised ANN based on the MLP Regressor. The network maps the 77 observed $H(z)$ values to the model parameters using a $77\!\to\!64\!\to\!32\!\to\!3$ architecture with ReLU activation. The model was trained for 1000 iterations and uncertainties were estimated from an ensemble of six independently trained networks. Training samples were generated from the theoretical $H(z)$ model using uniform priors $H_0 \in [65,75]\ \mathrm{km\,s^{-1}\,Mpc^{-1}}$, \; $\lambda \in [0.42,0.90]$, and $\omega_0 \in [-0.40,0.40]$ and 2000 noisy realizations consistent with OHD errors. The final posterior constraints is obtained from the ensemble predictions with 5000 Monte Carlo samples.
\begin{figure}
	\centering
	\begin{subfigure}[b]{0.7\textwidth}
		\centering
		\includegraphics[width=\textwidth]{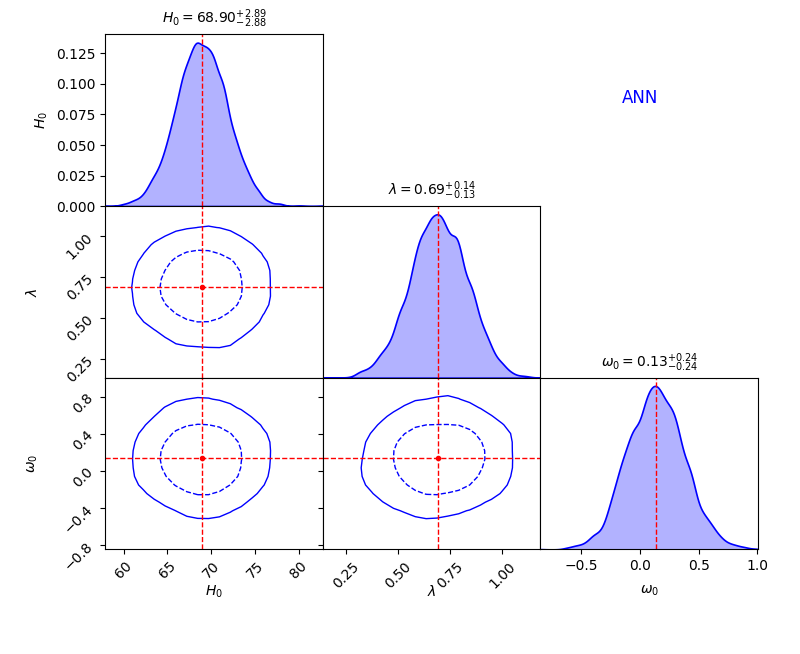}
		\caption{$H_{0}$, $\lambda$, and $\omega_{0}$ $1\sigma$ and $2\sigma$ contours from 77 $H(z)$ data points using ANN.}
	\end{subfigure}
	\vfill
	\begin{subfigure}[b]{0.8\textwidth}
		\centering
		\includegraphics[width=\textwidth]{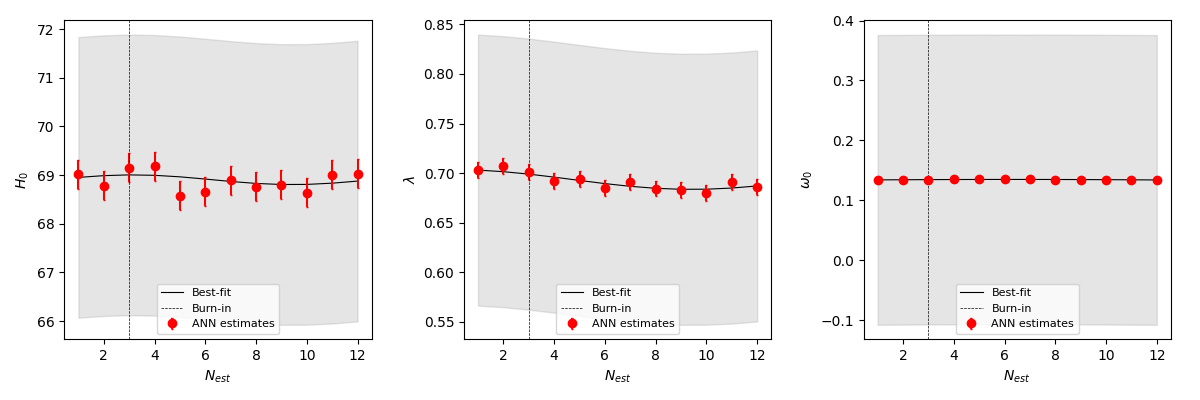}
		\caption{The relationship between steps and best-fit values and  $1 \sigma$ error of model parameters using ANN technique. The solid black line and Grey-shaded sections display the best-fit values, red circle with error bars are the results estimated by ANN Approach. }
	\end{subfigure}
	\vfill
	\begin{subfigure}[b]{0.5\textwidth}
		\centering
		\includegraphics[width=\textwidth]{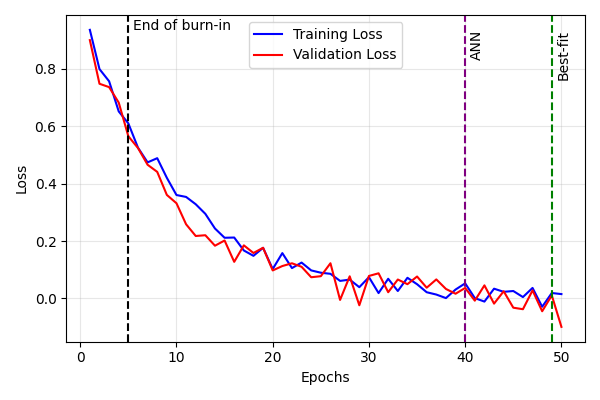}
		\caption{Plot of Losses of the training and validation sets. The training set consists of 3000 samples, while validation set contains 500 samples.}
	\end{subfigure}
	\caption{Observational analysis using ANN model}
\end{figure}
\subsection{Artificial Neural Networks}
\noindent In Ref. \cite{NIPS2017_addfa9b7}, the loss function for ANN method is read as
\begin{equation}
\mathcal{L} = \mathbb{E} \left( \frac{1}{N} \sum_{i=1}^{N} \left| \theta_i - \hat{\theta}_i \right| \right),
\end{equation}
The parameter space point \( \theta \) represents the estimated parameters, while the symbol \( N \) denotes the number of cosmological parameters and the target is symbolized by \( \hat{\theta} \) in the training set. 
\subsection{Mixture Density Network}
\noindent The loss function of MDNs methods is read as
\begin{equation}
\mathcal{L} = \mathbb{E} \left[ -\ln \left( \sum_{i=1}^{K} \omega_i \cdot \frac{\exp \left( -\frac{1}{2} (\hat{\theta} - \mu_i)^\top \Sigma_i^{-1} (\hat{\theta} - \mu_i) \right)}{\sqrt{(2\pi)^N |\Sigma_i|}} \right) \right], \label{eqn:loss_multi}
\end{equation}
where $\hat{\theta}$ (or $\hat{\boldsymbol{\theta}}$) represents the estimated parameters in both the single- and multi-parameter cases. 
\subsection{Mixture Neural Network}
\noindent The modified loss function for model with multiple parameters is obtained as
\begin{equation}
\mathcal{L} = \mathbb{E} \left[ - \ln \left( \sum_{i=1}^{K} \omega_i \cdot \frac{\exp \left( -\frac{1}{2} (\boldsymbol{\theta} - \hat{\boldsymbol{\theta}})^\top \boldsymbol{\Sigma}_i^{-1} (\boldsymbol{\theta} - \hat{\boldsymbol{\theta}}) \right)}{(2\pi)^{n/2} |\boldsymbol{\Sigma}_i|^{1/2}} \right) \right], \label{eqn:loss_multi}
\end{equation}

\noindent In this setup, the cosmological parameters $\theta$ (or $\boldsymbol{\theta}$) are considered as specific points within the parameter space. The target parameter values, $\hat{\theta}$ (or $\hat{\boldsymbol{\theta}}$), are derived from the training set, while $\boldsymbol{\omega}$ represents the mixture weights. \\

\begin{figure}
	\centering
	\begin{subfigure}[b]{0.7\textwidth}
		\centering
		\includegraphics[width=\textwidth]{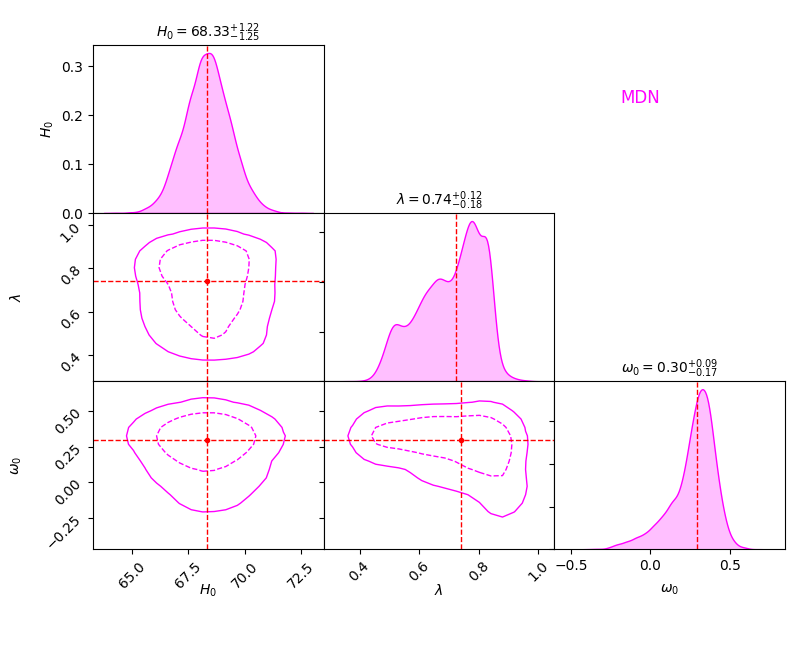}
		\caption{$H_{0}$, $\lambda$, and $\omega_{0}$ $1\sigma$ and $2\sigma$ contours from 77 $H(z)$ data points using MDN.}
	\end{subfigure}
	\vfill
	\begin{subfigure}[b]{0.8\textwidth}
		\centering
		\includegraphics[width=\textwidth]{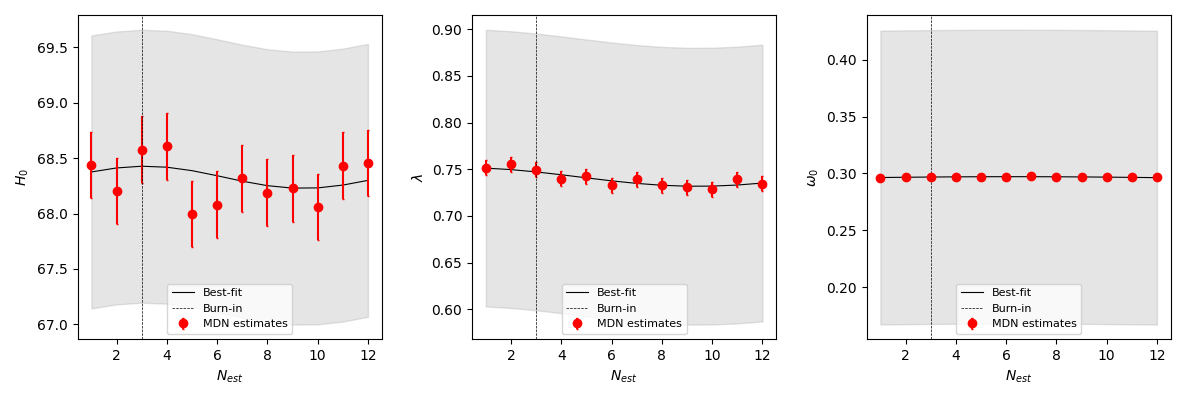}
		\caption{The relationship between steps and best-fit values and  $1 \sigma$ error of model parameters using MDN technique. The solid black line and Grey-shaded sections display the best-fit values, red circle with error bars are the results estimated by MDN Approach. }
	\end{subfigure}
	\vfill
	\begin{subfigure}[b]{0.5\textwidth}
		\centering
		\includegraphics[width=\textwidth]{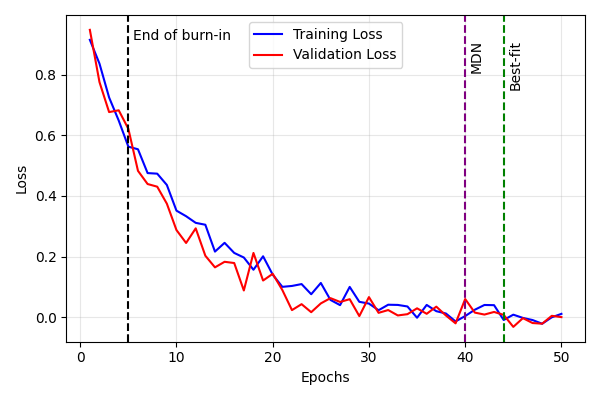}
		\caption{Plot of Losses of the training and validation sets. The training set consists of 3000 samples, while validation set contains 500 samples.}
	\end{subfigure}
	\caption{Observational analysis using MDN model}
\end{figure}


\begin{figure}
	\centering
	\begin{subfigure}[b]{0.7\textwidth}
		\centering
		\includegraphics[width=\textwidth]{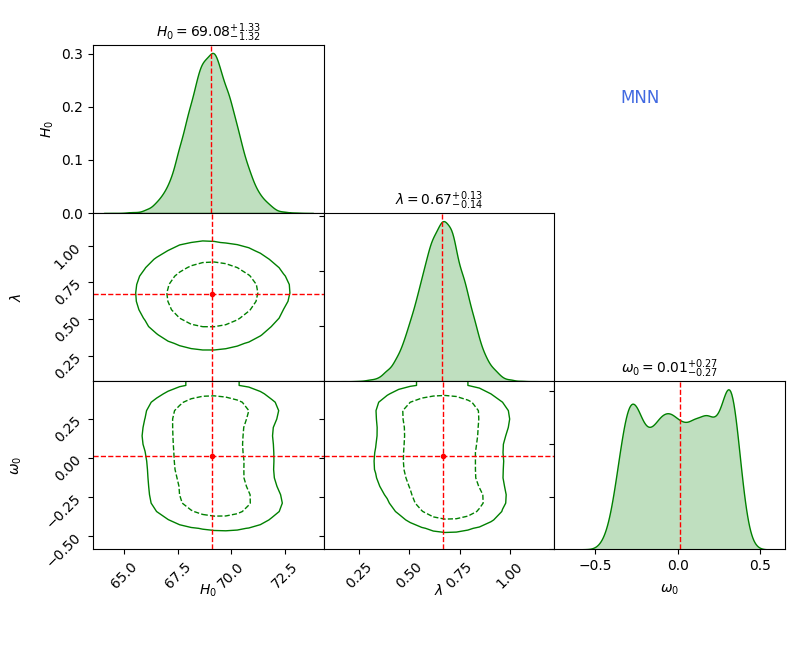}
		\caption{$H_{0}$, $\lambda$, and $\omega_{0}$ $1\sigma$ and $2\sigma$ contours from 77 $H(z)$ data points using MNN.}
	\end{subfigure}
	\vfill
	\begin{subfigure}[b]{0.8\textwidth}
		\centering
		\includegraphics[width=\textwidth]{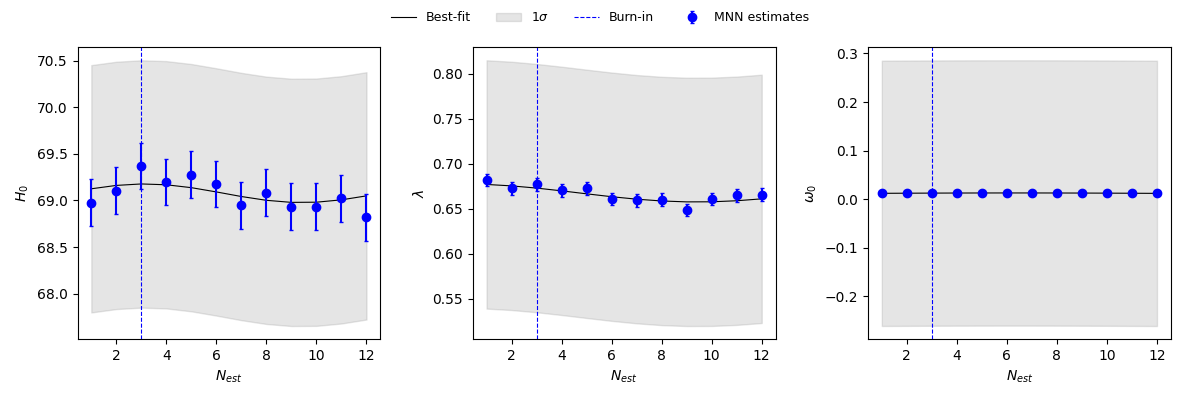}
		\caption{The relationship between steps and best-fit values and  $1 \sigma$ error of model parameters using MNN technique. The solid black line and Grey-shaded sections display the best-fit values, red circle with error bars are the results estimated by MNN Approach. }
	\end{subfigure}
	\vfill
	\begin{subfigure}[b]{0.5\textwidth}
		\centering
		\includegraphics[width=\textwidth]{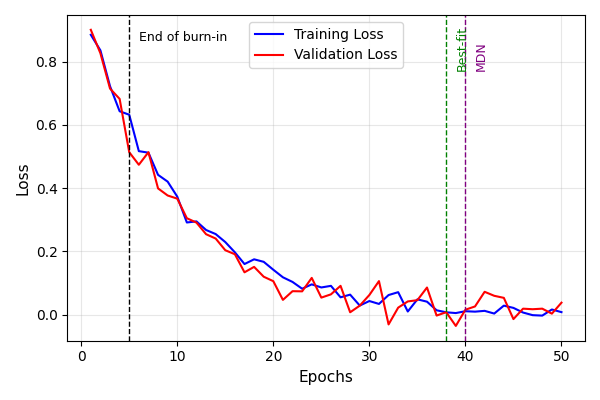}
		\caption{Plot of Losses of the training and validation sets. The training set consists of 3000 samples, while validation set contains 500 samples.}
	\end{subfigure}
	\caption{Observational analysis using MNN model}
\end{figure}


\section{Examination of the model's cosmological features}
\noindent For the derived model in Rastall gravity, the expressions for energy density and cosmic pressure are obtained as 
\begin{equation}
\rho = \frac{3 H_0^2 (4 \lambda -1) (z+1) \bigg[(z+1)^{3 \lambda -1} \left(\frac{(3 \lambda -1) z}{3 \lambda  \omega _0+3 \lambda -1}+1\right)\bigg]^{\frac{1-4 \lambda }{\lambda -3 \lambda ^2}}}{3 \lambda  \omega _0+(3 \lambda -1) (z+1)}
\end{equation}
\begin{equation}
p =  \frac{3 H_0^2 (4 \lambda -1) \omega _0 \bigg[(z+1)^{3 \lambda -1} \left(\frac{(3 \lambda -1) z}{3 \lambda  \omega _0+3 \lambda -1}+1\right)\bigg]^{\frac{1-4 \lambda }{\lambda -3 \lambda ^2}}}{3 \lambda  \omega _0+(3 \lambda -1) (z+1)}
\end{equation}


\subsection{Deceleration parameter}
The deceleration parameter (DP) form Hubble parameter expression can be formulated as $q= -1 + \frac{(1+z)}{H(z)} \frac{d H(z)}{dz} $. Thus, for the suggested model DP $q$ can be derived as:
\begin{equation}
q =  \frac{(6 \lambda -3) \omega_0+(6 \lambda -1) (z+1)}{6 \lambda  \omega_0+2 (3 \lambda -1) (z+1)}
\end{equation}
\begin{figure}
	\centering
	\includegraphics[scale=0.5]{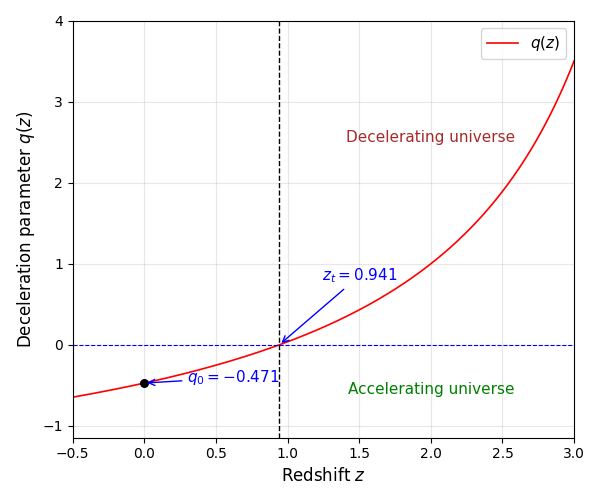}
	\caption{Plot of Deceleration parameter}
\end{figure}
In the present analysis, the model depicts a flipping nature of the universe from a deceleration era to a current expansion scenario that indicates a dark energy (DE) dominance in the present while a matter dominance in the past. The proposed model shows a transition behavior with the present value of DP as $q_0 = -0.471$ and signature flipping occurring at $z_{t} = 0.941$, as shown in Figure 7. The results obtained from the model align well with the latest experimental findings \cite{ref76,ref89,ref90,ref91}. Furthermore, it is important to note that the transition redshift $z_t=0.941$ is higher than the typical $\Lambda$CDM expectation $z_t\approx0.6$ - $0.8$ \cite{ref92,Planck2018}. This earlier transition can be interpreted as a natural consequence of the non--conservative matter sector in Rastall gravity \cite{ref36}, where the modified continuity equation permits an effective energy exchange between geometry and matter, altering the evolution of the effective dark--energy density. The relatively large value of the Rastall parameter ($\lambda\sim0.7$ from the ML reconstruction) strengthens this coupling and shifts the deceleration to acceleration transition to higher redshift. Hence, $z_t=0.941$ should be viewed as a characteristic prediction of non conservative cosmology rather than a tension with $\Lambda$CDM. Future precise measurements of the expansion history at intermediate redshifts will be important for testing this scenario.
\subsection{Statefinders diagnostic}
\begin{figure}
\centering
(a)\includegraphics[width=6.50cm,height=5.0cm,angle=0]{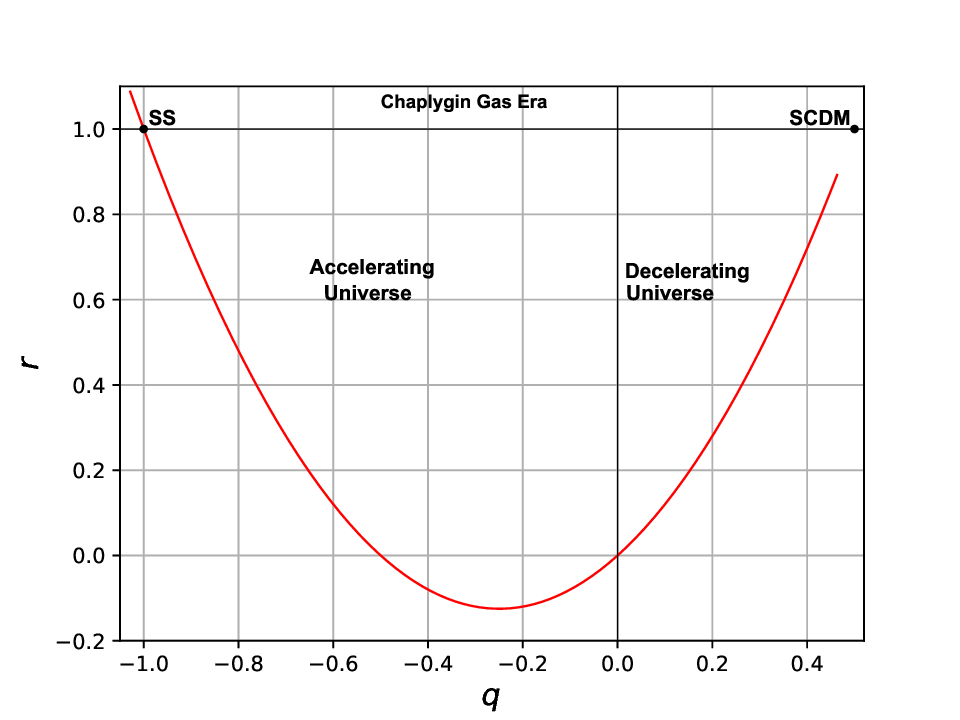}
(b)\includegraphics[width=6.50cm,height=5.0cm,angle=0]{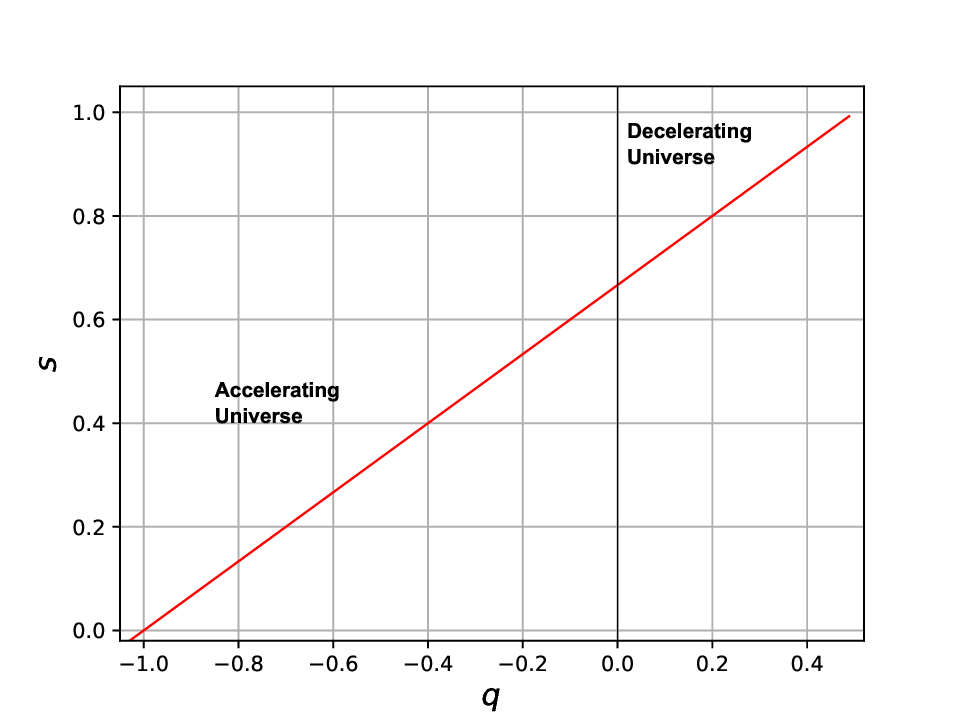}
(c)\includegraphics[width=6.50cm,height=5.0cm,angle=0]{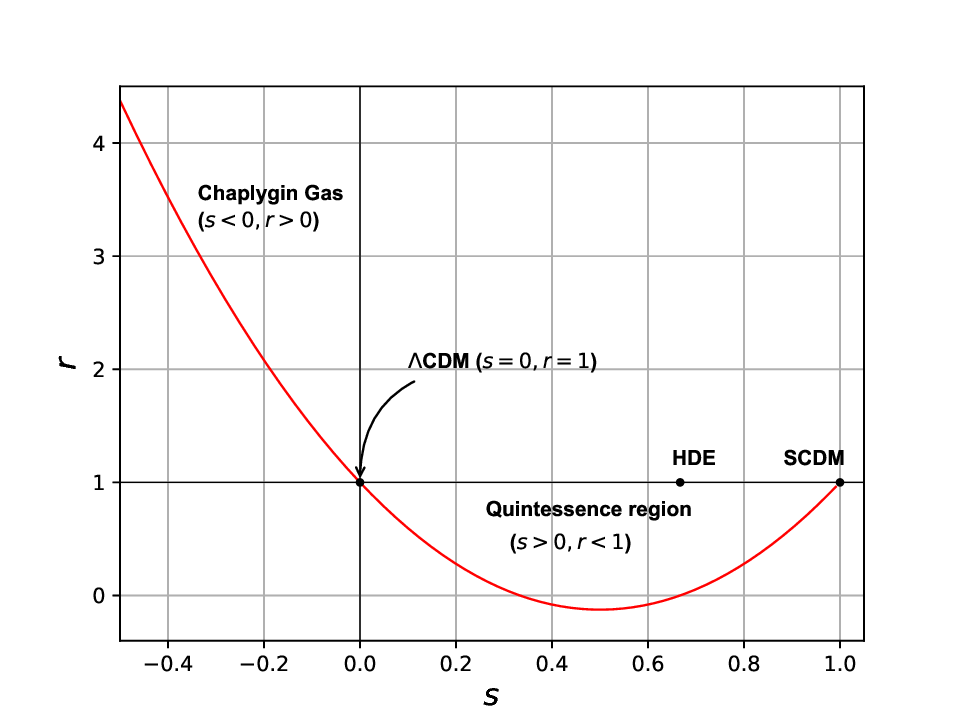}
\caption{(a) Plot of r vs q, (b) plot of s vs q, (c) Plot of r vs s.}
\end{figure}

Two novel parameters $r$ and $s$, named as statefinders, are introduced to distinguish the various dark energy cosmological models. The statefinder pair assists in improving the precision of model predictions by pinpointing the evolutionary path in the $r-s$ plane. By considering various forms of dark energy, as mentioned in the literature \cite{ref92a,ref93,ref94,ref95}, the distinction between the proposed cosmological model and the $\Lambda CDM$ model can be clearly distinguished on the $(r - ~s)$ plane \cite{ref92a,ref93}. For the derived model, the parameters $r$ and $s$ in the standard forms can be derived as:
\begin{eqnarray}
r&=&2 q^{2}+q - \frac{\dot{q}}{H}\nonumber\\
&=& \frac{9 \left(6 \lambda ^2-5 \lambda +1\right) \omega_0^2+6 \left(18 \lambda ^2-9 \lambda +1\right) \omega_0 (z+1)+\left(54 \lambda^2-21 \lambda +2\right) (z+1)^2}{2 \left[3 \lambda  \omega_0+(3 \lambda -1) (z+1)\right]^2}
\end{eqnarray}

\begin{eqnarray}
s&=&\frac{r-1}{3 (q-\frac{1}{2})}\nonumber\\
&=& \frac{(4 \lambda -1) \left[3 (\lambda -1) \omega _0^2+2 (3 \lambda -1) \omega_0 (z+1)+3 \lambda  (z+1)^2\right]}{3 \left((\lambda -1) \omega_0+\lambda  (z+1)\right) \left(3 \lambda  \omega _0+(3 \lambda -1) (z+1)\right)}
\end{eqnarray}
Figure 8 illustrates the features of the suggested model in the $r-s$ plane using formulas for $r$ and $s$. The plot shows that the suggested model is situated within the Chaplygin gas region $(r > 1, s < 0)$ and converges to the $\Lambda$CDM point $(r = 1, s = 0)$ at the late passes of the timeline. For the present epoch, the statefinder parameters ($r, s$) are obtained as (0.035, 0.312). Hence, the derived model shows features of an accelerating cosmic model other than $\Lambda$CDM \cite{ref92a,ref94,ref95,ref97}.  

\noindent

\subsection{Jerk Parameter}

The jerk parameter ($j$) serves as a widely utilized tool in the field of cosmology to diagnose the various cosmological models. The concept arises from the belief that there must be a sudden jolt to shift the universe from a decelerating to an accelerating phase. The jerk is defined as the acceleration's rate of change over time from a physical perspective. It originates from the third-order term of Taylor's series expansion of the scale factor $a$ centered at $a_0$ in cosmology. This parameter helps to differentiate between cosmological models that are kinematically degenerate. The involvement of the third order derivative of the scale factor results in increased accuracy when describing the expansion of the universe compared to the Hubble parameter.  The parameter $j$ for the suggested model can be expressed as \cite{ref97a}:
\begin{eqnarray}
j&=&1-(1+z) \frac{H'(z)}{H(z)}+\frac{1}{2}(1+z)^2 \left[\frac{H''(z)}{H(z)}\right]^2
\end{eqnarray} 
Here $H'(z)$ and $H''(z)$ respectively denote the first and second order derivatives of parameter $H(z)$ w.r.to redshift $z$. For the given model parameter $j$ can be recast as:

\begin{eqnarray}
j &=& -\frac{9 (1-4 \lambda )^2 \left[(6 \lambda -3) \omega _0^2+2 (6 \lambda -1) \omega _0 (z+1)+(6 \lambda -1) (z+1)^2\right]^2}{32 (z+1)^2 \left[3 \lambda  \omega _0+(3 \lambda -1) (z+1)\right]^4}\nonumber\\
&+&\frac{(3-6 \lambda ) \omega _0-(6 \lambda -1) (z+1)}{6 \lambda  \omega _0+2 (3 \lambda -1) (z+1)}
\end{eqnarray} 
\begin{figure}
	\centering
	\includegraphics[scale=0.5]{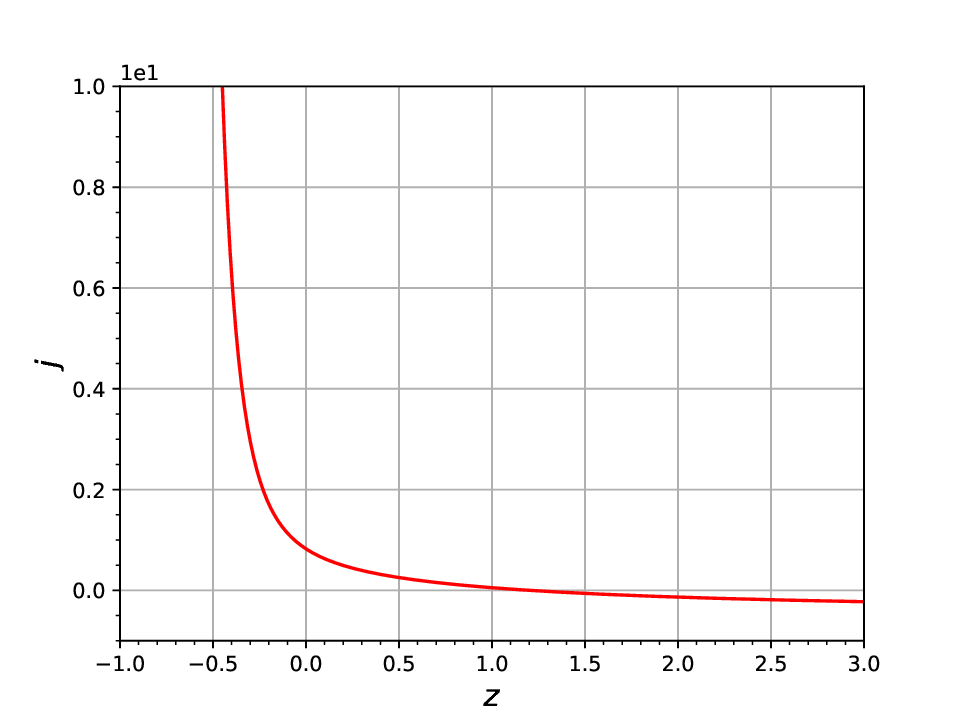}
	\caption{Behaviour of jerk parameter $ j $ with $ z $. }	
\end{figure}

Figure 9 illustrates the evolution of the jerk parameter $j$ as a function of redshift. Earlier theoretical studies suggest that positive values of the jerk parameter indicate an accelerated expansion of the universe \cite{ref97b,ref97c,ref97d}. 
Based on our model, the jerk parameter is initially negative and then increases, eventually convergent to positive over time. In the existing analysis, the present value of the parameter $j_0$ is determined to be $0.824$ for the joint observational dataset.

\subsection{Om diagnostic}
The parameter $ O_{m}$ serves as an additional diagnostic tool in cosmology, as introduced by Sahni et al. \cite{ref97e}. This parameter can be derived solely from the Hubble parameter, eliminating the need for $H'(z)$ or other pertinent data, which consequently reduces the likelihood of errors. Its ability to allow for reconstruction through both non-parametric as well as parametric approaches has contributed to the popularity of $O_{m}$ in cosmological investigations. Furthermore, it is important to mention that the $O_{m}$ can differentiate between various cosmic models even in the absence of information regarding the matter density parameter and the equation of state (EoS) \cite{ref97e,ref97f}. According to the standard formulation \cite{ref97e}, the expression for $ O_{m} (z) $ in the given model can be expressed as follows:
\begin{equation}
Om(z)=\frac{\left((z+1)^{3 \lambda -1} \left(\frac{(3 \lambda -1) z}{3 \lambda  \omega _0+3 \lambda -1}+1\right)\right){}^{\frac{1-4 \lambda }{\lambda -3 \lambda ^2}}-1}{z^3+3 z^2+3 z}
\end{equation}
\begin{figure}
\centering
\includegraphics[scale=0.5]{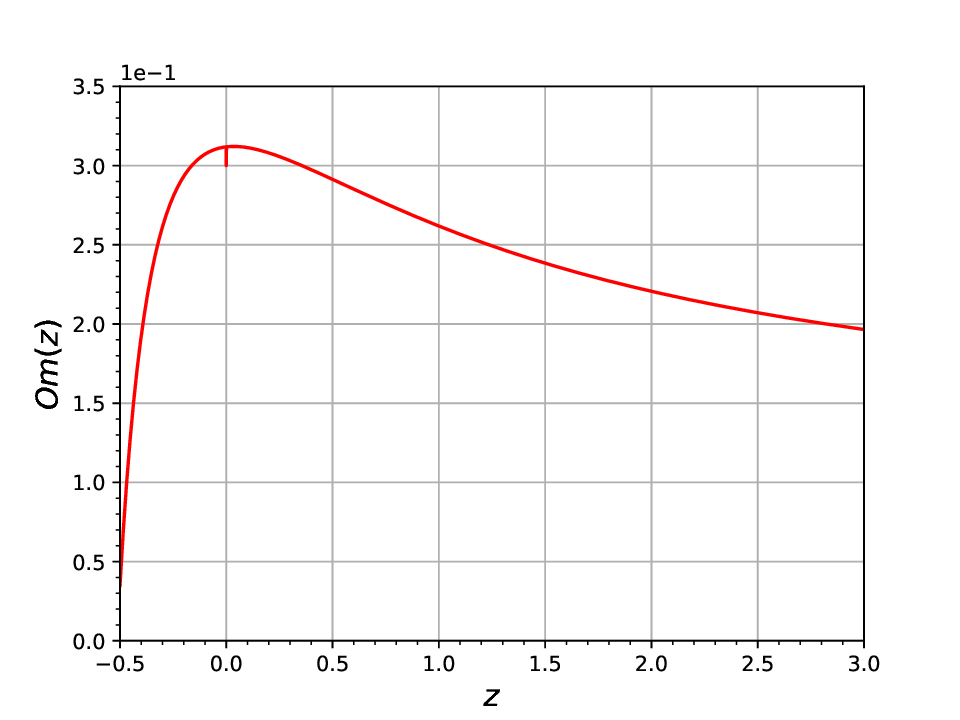}
\caption{Evaluation of $ Om(z) $.}	
\end{figure}
The evolution trajectory of $O_{m} (z)$ has been displayed in Figure 10. The incline of the diagnostic $O_{m}(z)$ serves as a crucial indicator for determining the type of dark energy (DE) cosmic models. Specifically, a negative slope suggests a quintessence-like model, while a positive slope indicates a phantom-like cosmic model. Furthermore, a zero curvature in the slope of $O_{m}(z)$ implies a $\Lambda$CDM type of DE model. The $O_{m}(z)$ analysis of our model reveals a positive slope, indicating a phantom-like scenario that supports the belief of a dark energy-dominated universe. The suggested cosmological model is distinguished from the standard $\Lambda$CDM model by its positive curvature \cite{ref34,ref97b,ref97f}.
\section{Energy Conditions}
We have discussed the physical viability of the derived model via the evolution of energy conditions (ECs) \cite{ref98,ref99}. The energy conditions for cosmic model in Rastall theory are defined as\cite{ref98,ref99,ref100}: (i) Weak energy conditions (WEC) if $\rho \geq0$, (ii) Dominant energy conditions (DEC) if $\rho - p \geq 0$, (iii) Null energy conditions (NEC) if $\rho + p \geq 0$, and (iv) Strong energy conditions (SEC) if $\rho + 3 p \geq 0$. Generally, the WEC and DEC are always satisfied by all recognized energy and matters \cite{ref101,ref102,ref103}. The unusual energy (DE) that creates a strong negative pressure is responsible for the universe's fast expansion and the violet SEC \cite{ref101,ref102,ref103,ref104}. \\
\begin{figure}\label{ec}
\centering
\includegraphics[scale=0.5]{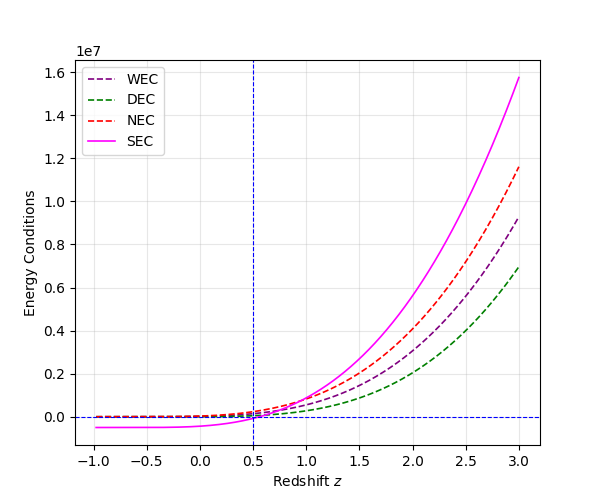}
\caption{Plot of Energy Conditions. }	
\end{figure}

\noindent For a perfect fluid in Rastall gravity, the strong energy condition (SEC) takes the modified form  
\begin{equation}
\rho + 3p - \frac{2\lambda}{4\lambda - 1}(\rho - 3p) \geq 0.
\end{equation}
The additional $\lambda$-dependent contribution effectively shifts the boundary for cosmic acceleration. Consequently, fluids that would ordinarily satisfy $\rho + 3p > 0$ in the standard (bare) sector may, for certain values of $\lambda$, lead to an effective violation of the SEC thereby permitting accelerated expansion without the need for exotic matter components. The coupling parameter $\lambda$ is subject to stringent constraints: (i) $\lambda \neq 1/4$ on theoretical grounds, and (ii) $0.27 \lesssim \lambda \lesssim 0.74$ from cosmological data fits. These features highlight how Rastall gravity allows energy-condition violations to arise naturally, but only within a relatively narrow and observationally restricted parameter space. \\

\noindent Table~2 summarizes the statistical comparison between the MCMC analysis and the machine learning approaches (ANN, MDN and MNN). Using the OHD dataset, the MCMC method yields $H_{0}=70.53^{+1.01}_{-1.02}\,\mathrm{km\,s^{-1}\,Mpc^{-1}}$, $\omega_{0}=1.00^{+0.03}_{-0.03}$ and $\lambda=0.28^{+0.00}_{-0.00}$, indicating a relatively rigid cosmological behaviour. In contrast, the ML methods consistently predict $H_{0}\approx 68$ - $69\,\mathrm{km\,s^{-1}\,Mpc^{-1}}$, with lower values of the equation of state parameter $\omega_{0}\sim 0.01$ - $0.30$ and significantly higher Rastall parameter $\lambda\approx 0.67$ - $0.74$. These results indicate improved parameter clustering and a more flexible reconstruction of the cosmological dynamics using ML techniques.\\

\noindent Figure 11 shows the behavior of various energy conditions in Rastall gravity at present epoch. In this analysis, the Dominant Energy Condition (DEC) and Null Energy Condition (NEC) are satisfied while the Strong Energy Condition (SEC) violates for the proposed model as depicted in Figure that validates an accelerated expansion of cosmos in the present era \cite{ref101,ref102,ref103,ref104,ref105}.\\
\begin{table}
\begin{center}
\begin{tabular}{|c|c|c|c|}
\hline 
Parameters &  $H_{0}$ &  $\omega_{0}$	& $\lambda$   \\
\hline
OHD &  $70.53^{+1.01}_{-1.02}$ &  $1.00^{+0.03}_{-0.03}$  &  $0.28^{+0.00}_{-0.00}$  \\ 
\hline
BAO &  $69.0^ {+1.41}_{-1.43}$ &  $1.00^{0.08}_{-0.08}$  &  $0.27^{0.00}_{0.00}$  \\ 
\hline
Pantheon+ &  $72.55^{+0.29}_{-0.28} $ &  $ 0.97^{+0.30}_{-0.30} $  & $0.29^{+0.01}_{-0.02}$  \\ 
\hline
ANN &  $68.90^{+2.89}_{-2.88} $ &  $0.13^{+0.24}_{-0.24}$  &  $0.69^{+0.14}_{-0.13}$  \\
\hline
MDN &  $68.33^{+1.22}_{-1.25} $ & $0.30^{+0.09}_{-0.17}$  &  $0.74^{+0.12}_{-0.18}$ \\
\hline
MNN &  $69.08^{+1.33}_{-1.32} $ &  $0.01^{+0.27}_{-0.27}$  &  $0.67^{+0.13}_{-0.14}$ \\
\hline
\end{tabular}\caption{ The best-fit values of model parameters for the proposed model }
\end{center}
\end{table}
\section{Information criteria and model selection}
\noindent  
As a final step, we apply the known Akaike Information Criterion (AIC) \cite{Akaike1974stat} and the Bayesian Information Criterion (BIC) \cite{Schwarz1978model}, in order to examine the quality of the fittings and hence the relevant observational compatibility of the scenarios. The AIC is based on information theory, and it is an estimator of the Kullback-Leibler information with the property of asymptotically unbiasedness. Under the standard assumption of Gaussian errors, the corresponding estimator reads as \cite{Anderson2002model,Burnham2004inference}
\begin{equation*}
AIC  = -2 \ln ({\mathcal{L}_{max}})+2 k +\frac{2 k (k+1)}{N-k-1}
\end{equation*}	
here, $k$ denotes the number of free variables in model, $\ln {\mathcal{L}_{max}} $ the maximum likelihood of the datasets and N is the total data points. For large number of data points, it reduces to $AIC  -2 \ln ({\mathcal{L}_{max}})+2 k$. On the other hand, the BIC criterion is an estimator of the Bayesian evidence \cite{Anderson2002model,Burnham2004inference,Liddle2007information}, given by
\begin{equation*}
BIC  = -2 \ln ({\mathcal{L}_{max}})+2 k \ln N
\end{equation*}	
To be more precise, a model having $0 \leq \Delta AIC < 2$ and $0 \leq \Delta BIC < 2$ receives strong evidence in favor. In contrast, for $2 < \Delta AIC < 4$ and for $2 \leq \Delta BIC < 6$, the model has average evidence in favor, whereas for $4 < \Delta AIC < 7$ and $6 \leq \Delta BIC < 10$, the model is considered to have less evidence in favor; and, finally, for $ \Delta AIC > 10$ or $ \Delta BIC > 10$, the model receives no significant support since it has no evidence in favor \cite{Liddle2007information}.\\

\noindent Our analysis shows that the $\Lambda$CDM model is preferred by the $\Delta$AIC and $\Delta$BIC criterion. For OHD data sets, we immediately find that $\Delta AIC = 2.43$ and $\Delta BIC = 3.44$. Since, this model shows the differences in the range of $ 4 < \Delta AIC < 7$ and $ 2 \leq \Delta BIC < 6$, therefore, this model has average evidence in favor. \\


\section{Conclusion}
\noindent In this study, we have explored the cosmological dynamics of an isotropic, homogeneous universe in Rastall gravity. To determine the explicit solution for field equations in Rastall gravity, the parameterization of EoS parameter in the form $\omega(z) = \frac{\omega_{0}}{(z+1)} $ is considered, and the Hubble parameter as a redshift function is derived. The model parameters are estimated by taking observational BAO, Pantheon plus compilation of SN Ia, and OHD datasets using MCMC analysis. To constrain the cosmological parameters and analyze the current cosmic scenario of the accelerating universe, an innovative deep-learning approach is also employed simultaneously. Using the CoLFI Python package and advanced techniques such as Artificial Neural Networks (ANN), Mixture Density Networks (MDN), and Mixture of Gaussians (MNN), the model parameters are estimated efficiently. This advancement significantly facilitated the understanding of conditional probability densities derived from observational data and posterior distributions. To enhance the efficiency of neural network training and improve the accuracy of parameter predictions, hyperellipsoid parameters were introduced. A comparative study between our neural network methods and traditional Markov Chain Monte Carlo (MCMC) techniques revealed that ANN produced nearly identical results, confirming its reliability and effectiveness.\\  

\noindent The proposed model of the universe efficiently explains the late-time acceleration without the need for dark energy and avoids issues associated with the cosmological constant. Therefore, it is essential to consider current observational data when simulating Rastall gravity. We have shown that neural network methods can effectively estimate cosmological parameters, serving as a practical alternative to MCMC techniques. The Rastall gravity framework facilitates the exploration of various gravity theories and their influence on cosmic expansion through the use of ANN, MDN, and MNN methodologies. This research is part of a broader study integrating machine learning with cosmology, demonstrating the capability of machine learning to address intricate cosmic dynamics and expansion challenges. The analysis of statefinders indicates that the resulting model exhibits characteristics of an accelerating cosmic model other than $\Lambda$CDM. The analysis of the jerk parameter $j(z)$ and $O_m (z)$ diagnostic for the derived model shows the phantom-like behavior. The violation of the Strong Energy Condition (SEC) for the proposed model validates an accelerated expansion of the cosmos in the present era. Moreover, the AIC/BIC analysis clearly favors the $\Lambda$CDM model over the present scenario. Using the OHD dataset, we obtained $\Delta AIC = 2.43$ and $\Delta BIC = 3.44$ which correspond to average evidence against the proposed model and in favor of $\Lambda$CDM. Therefore, although the model provides an acceptable fit to the observational data and remains phenomenologically viable, the statistical model comparison indicates that $\Lambda$CDM is still preferred by current observations. \\
 

\noindent  Furthermore, deep learning techniques offer a distinct advantage over conventional machine learning (ML) approaches in feature extraction. Classical ML generally requires manual feature engineering, followed by prediction based on these handcrafted inputs, whereas deep learning models automatically learn hierarchical features directly from the data. In this work, we employ deep learning to predict cosmological parameters, consistent with recent applications in cosmology \cite{Schmelzle/2018,Ribli/2019,Fluri/2019,He/2019,Lucie-Smith/2020}.\\ 

\noindent In our framework, cosmological model parameters are initialized within specified intervals, and synthetic instances are generated to construct training and validation datasets. Observational uncertainties and random noise are incorporated into the training set to enhance robustness. After preprocessing, an ANN is built according to the dataset size and trained for 2000 epochs. The observational data are then passed through the network to infer the cosmological parameters. \\

\section*{Acknowledgement}
\noindent The authors are very grateful to the honorable referee for constructive suggestions which have significantly improved our work in terms of research quality and presentation.

\end{document}